\documentclass[smallextended]{svjour3} 
\usepackage{bm}
\usepackage{graphicx} 
\usepackage{array} 
\usepackage{verbatim} 
\usepackage{subfig} 
\usepackage{tikz} 
\usetikzlibrary{shapes,snakes}
\usetikzlibrary{backgrounds,fit,decorations.pathreplacing} 
\newcommand{\ket}[1]{\ensuremath{\left|#1\right\rangle}} 
\usepackage{multirow}
\usepackage{marvosym}
\newcommand{\envelope}{(\raisebox{-.5pt}{\scalebox{1.45}{\Letter}}\kern-1.7pt)}
\usepackage{amsmath}
\usepackage{afterpage}
\usepackage{rotating}
\usepackage[nottoc,notlof,notlot]{tocbibind} 
\usepackage[titles,subfigure]{tocloft} 

\usepackage{color}
\definecolor{darkgreen}{rgb}{0,0.4,0}
\definecolor{darkblue}{rgb}{0,0,0.4}

\begin{document}
\title{A Universal Quantum Circuit Scheme For Finding Complex Eigenvalues}
\author{Anmer~Daskin \and Ananth~Grama \and Sabre~Kais
}
\institute{A. Daskin -A. Grama \at
Department of Computer Science, Purdue University, West Lafayette, IN, 47907 USA
\and
S. Kais \at
Department of Chemistry, Department of Physics and Birck Nanotechnology Center,Purdue University,
West Lafayette, IN 47907 USA;
Qatar Environment and Energy Research Institute, Doha, Qatar
}
\date{Received: date / Accepted: date}
\maketitle

\begin{abstract}
We present a general quantum circuit design for finding eigenvalues of non-unitary
matrices on quantum computers using the iterative phase estimation algorithm. In addition, 
we show how  the method can be used for the simulation of resonance states for quantum systems.  
\end{abstract}

\section{Introduction}
In the circuit model of quantum computation, a controlled quantum system, considered
to be a quantum computer, propagates from one state to another through the application
of a local  propagator called a quantum gate. A given algorithm, or computation, can
be  implemented on a quantum computer using  a sequence of quantum gates. These gates
can be represented by unitary matrices in the computational basis. While this unitary
representation is adequate for many problems, it impedes application of quantum computing
to new  problems, where computations can only be described through non-unitary
matrices. Furthermore, any attempt at approximation of these non-unitary parts negatively
impacts the accuracy of the results. For instance, the algorithm of Harrow et. al \cite{Lloyd}
for solving linear systems of equations requires a non-unitary quantum gate. The
circuit designed for this algorithm \cite{Yudong} uses an approximated gate for the
non-unitary  parts.

Quantum algorithms are known to be more efficient than their classical
counterparts \cite{Deutsch,Bennett,Delgado}. They often provide exponential efficiency
gains for the simulation of quantum systems \cite{Abrams,Papageorgiou}. 
The idea of  making a computer compatible, from the start, with the principles
of quantum theory for simulations of quantum systems was discussed in
1969 by Finkelstein \cite{Finkelstein}. Later in 1981, Feynman \cite{Feynman} explained
why it is impossible to simulate a quantum system using a probabilistic classical
computer. He also described a universal quantum computer capable of simulating quantum
physics. Since then, different algorithms and quantum circuit designs have been proposed
for the simulation of various quantum systems -- simulation of chemical
dynamics \cite{Sanders2009algorithm,Sanders,Kassal2008polynomial}, simulation of sparse
Hamiltonians \cite{Berry2007}, calculating the thermal rate constant \cite{New3},
and others \cite{Brown2010,New6,Kassal2011,Whaley2012}. Arguably, the most important
one of these algorithms is the phase estimation algorithm \cite{Abrams,Kitaev},
used for finding the eigenvalues of a given unitary matrix, or equivalently, the
eigen-energies of the corresponding quantum system. The algorithm has been applied in quantum
chemistry to obtain energies of molecular systems \cite{New1,New2,New4,Daskin2}.
The algorithm has also been demonstrated experimentally for the simulation of the
hydrogen molecule using photonic \cite{New5,Alan2012} and NMR quantum
computers \cite{New6,Du2010}. One of the postulates of quantum mechanics states
that measurable quantities such as  the energy of stable atoms and molecules are
real quantities and the operator that represents them should be a Hermitian operator.
Because the eigenvalues of Hermitian operators are real, the expectation of any
measurable quantity related to these eigenvalues is also real. However, it is
almost impossible to obtain the poles of the scattering matrix within the framework
of the standard formalism of the quantum mechanics, where we use only Hermitian operators.
In the non-Hermitian formalism, the poles of the scattering matrix can be directly
calculated from the complex eigenvalues of the non-Hermitian Hamiltonian \cite{Moiseyev2011}.
The simulation of non-Hermitian matrices in quantum theory requires the simulation
of non-unitary evolution matrices. However, quantum computers are based on the standard
formalism; hence, computations are done using unitary evolution operators. 

It has been shown that a quantum circuit can be generalized to a non-unitary circuit,
whose constituents are non-unitary gates representing quantum measurement. Furthermore, it is
shown that a specific type of one-qubit non-unitary gates, the controlled-NOT gate,
and all one-qubit unitary gates constitute a universal set of gates for the non-unitary
quantum circuit without the necessity of introducing ancilla qubits \cite{Terashima}. 
Recently, Wang et al. \cite{Hefeng-MPEA}, proposed a measurement-based quantum algorithm
for finding eigenvalues of non-unitary matrices. Their method draws on ideas from
conventional phase estimation algorithm, frequent measurement, and techniques in
one-qubit state tomography. They  describe a bipartite system composed of two subsystems,
where the total Hamiltonian includes the Hamiltonians of each subsystem and their
interaction. They show that different non-unitary evolution matrices can be constructed
by performing sequential projective measurements on one of these subsystem. In addition,
they show that an eigenvalue of the constructed  matrix can be estimated within the
phase estimation algorithm in two different ways: using the state tomography of a qubit
or the measured quantum Fourier transform with projective measurements to separately
compute the real and imaginary parts of the eigenvalue of the Hamiltonian. However,
the measurement described in their paper is a non-unitary process and, for this reason,
cannot be implemented deterministically. This is also an obstacle to directly applying
their method for a given non-unitary matrix without knowing the implementation of the
subsystems forming the corresponding Hamiltonian matrix. Furthermore, the process of
successive projective measurements changes the state to a greater extent
\cite{Terashima2012}, which effects the accuracy of the constructed non-unitary matrices,
and hence the computed eigenvalues in the proposed method. The success probability of the
algorithm depends on the success probabilities of the successive projective measurements
in each step of the algorithm, which may decrease exponentially. The  sequential
projective measurements in every step of the phase estimation algorithm causes another
issue -- the efficiency of the algorithm becomes dependent on the efficiency of the
implementations of the projective measurements. 

In this paper, we introduce a systematic way of estimating the complex eigenvalues
of a general matrix using the standard iterative phase estimation algorithm with a programmable
circuit design \cite{Emmar2}. The bit values of the phase of the target eigenvalue
is determined from the outcome of the measurement on the phase qubit. Then, the
statistics of the outcomes of the measurements on the phase qubit are used to
determine the absolute value of the eigenvalue. Consequently, using the phase and
the absolute value of the eigenvalue, the  complex eigenvalue of the non-unitary
is determined. Because of the exact circuit design used for the simulation of a
given non-unitary matrix, our method produces very accurate results. \textbf{The success
probability of the algorithm depends on the number of qubits needed in the ancilla,
and decreases exponentially with size, in the case of dense matrices.}
Our method can  be used as a general circuit equivalent for any non-unitary
matrix.  We also present the application of our method to an example of non-Hermitian
Hamiltonians. Since the proposed circuit design is universal, it can be used to estimate
the eigenvalues of any given system.

In the following sections, we first explain and verify the universal circuit
described in ref.\cite{Emmar2}. We then explain how to use the circuit within the
iterative phase estimation algorithm. We discuss the algorithmic complexity and
implementation issues, and compare our method to the work of Wang et al. \cite{Hefeng-MPEA}.
Finally, we explain how to apply the method to the simulation of non-Hermitian quantum systems. 

\section{Universal  Circuit for Non-Unitary Matrices}

For an arbitrary $n$-qubit system represented by  a matrix $U$ of size $N$ (where $N=2^n$), 
it is known that the relationship between an arbitrary input, assumed to be $\ket{\alpha}$,
to the system  $U$ and the corresponding output, $\ket{\beta}$, can be defined as
$U\ket{\alpha}=\ket{\beta}$:
\begin{equation}
	\label{eqbm}
	U\ket{\alpha}=
	\left(\begin{matrix}
	u_{11}&u_{12}&\dots&u_{1N}\\
	u_{21}&u_{22}&\dots&u_{2N}\\
	\vdots&\vdots&\ddots&\vdots\\
	u_{N1}&u_{12}&\dots&u_{NN}\end{matrix}\right)\left(\begin{matrix}\alpha_{1}\\
	\alpha_2\\\vdots\\ \alpha_{N}\end{matrix}\right)=
	\left(\begin{matrix}\beta_{1}\\
	\beta_2\\\vdots\\ \beta_{N}\end{matrix}\right),
\end{equation}

\begin{figure}
	\centering
	\includegraphics[scale=0.5]{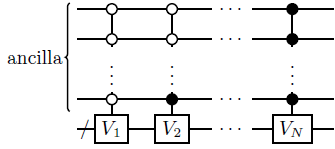}
	\caption{Each block operations forms a different row of the given matrix as their
leading row. \cite{Emmar2}}
	\label{fig:blocks}
\end{figure}

Recently \cite{Emmar2}, we have described a universal programmable circuit design
which can be used to simulate the action of  the matrix $U$ in Eq.(\ref{eqbm}) 
by simply determining the gate angles from the matrix elements.  
The  idea is to independently generate circuit equivalences for  the rows of $U$ 
in separate $N\times N$ block quantum operations; and then  combine the blocks 
 by using $2^{n+1}$ different states of $(n+1)$ ancillary control qubits. This is shown in
Fig.\ref{fig:blocks}. 
The circuit in Fig.\ref{fig:blocks} can  be represented by a block diagonal matrix $V$
with $N$ submatrices on the diagonal:
\begin{equation}
	V=
	\left(\begin{matrix}
		V_1&&&\\
		&V_2&&\\
		&&\ddots&\\
		&&&V_N
	\end{matrix}\right)
\end{equation}
 Here, each submatrix $V_i$  is to have the $i$th row of the given matrix 
as their first rows in the following form:
$\left[ku_{i1} \bullet ku_{i2} \bullet \dots ku_{iN} \bullet\ \right]$, \textbf{where
$k=1/\sqrt{N}$ is the normalization constant.} We use the sign \lq\lq{}$\bullet$\rq\rq{} for
the elements that are insignificant in the overall scheme. 

To use the matrix $V$ in place of the matrix $U$ in Eq.(\ref{eqbm}), we modify
the input $\ket{\alpha}$ in Eq.(\ref{eqbm}) to a vector $|\varphi\rangle$ in a
way that $|\varphi\rangle$ is populated with the elements of $\ket{\alpha}$.
Therefore, the expected output of $U\ket{\alpha}=\ket{\beta}$ can also be produced
on some predetermined  states in the outcome of $V|\varphi\rangle$. This is shown below:
\begin{equation}
	\label{eqbm2}
	V|\varphi\rangle=
	\left(\begin{matrix}
		\begin{matrix}
		ku_{11} &\bullet&\dots& ku_{1N}&\bullet\\
		\bullet&\bullet&\dots&\bullet&\bullet\\
		\vdots&\vdots&\vdots&\vdots&\vdots\\
		\bullet&\bullet&\dots&\bullet&\bullet\\
			\end{matrix}& &
		\\ 
		& \ddots &
		\\
		& &\begin{matrix}
		ku_{N1}&\bullet&\dots &ku_{NN}&\bullet	\\
		\bullet&\bullet&\dots&\bullet&\bullet\\
		\vdots&\vdots&\vdots&\vdots&\vdots\\
		\bullet&\bullet&\dots&\bullet&\bullet\\
		\end{matrix}
	\end{matrix}\right)
	\left(\begin{matrix}
		  \gamma\alpha_1\\
		  0\\
		    \gamma\alpha_2\\
		  \vdots\\
		    \gamma\alpha_N\\
		    0\\
		       \gamma\alpha_1\\
		         0\\
		    \gamma\alpha_2\\
		    \vdots\\
		 \gamma\alpha_N\\ 
		 0
	\end{matrix}\right)=
	\left(\begin{matrix}\kappa\beta_{1}\\ \bullet\\ \vdots\\ \bullet\\
		\kappa\beta_2\\ \bullet \\ \vdots\\ \bullet \\ \kappa\beta_{N}\\ 
		\bullet \\ \vdots\\ \bullet
		\end{matrix}\right),
\end{equation}
\textbf{where $\gamma=1/\sqrt{N}$ and $\kappa=1/N$ are the normalization constants.}
In the above equation,
the amplitudes \{$\kappa\beta_{1},\kappa\beta_{2},\dots,\kappa\beta_{N} $\} on the states
wherein the main $n$ qubits are zero are the same as the scaled expected outcomes of
the application of $U$ to $\ket{\alpha}$. The rest of the states in the output
represented by \lq\lq{}$\bullet$\rq\rq{} is to be ignored.
\textbf{Therefore, the probability of success 
of the post-selection on the zero state of the main qubits, which is the success
of the simulation, is given by $\kappa^2$.} As an example, if $N=2$, Eq.(\ref{eqbm2})
becomes as follows:
\begin{equation}
	\label{eqbm2explicit}
	V|\varphi\rangle=
	\frac{1}{\sqrt{2}}\left(\begin{matrix}
		\begin{matrix}
		u_{11} &\bullet& u_{12}&\bullet\\
		\bullet&\bullet&\bullet&\bullet\\
		\bullet&\bullet&\bullet&\bullet\\
		\bullet&\bullet&\bullet&\bullet\\
			\end{matrix}& &
		\\ 
		& &\begin{matrix}
		u_{21}&\bullet&u_{22}&\bullet	\\
		\bullet&\bullet&\bullet&\bullet\\
		\bullet&\bullet&\bullet&\bullet\\
		\bullet&\bullet&\bullet&\bullet\\
		\end{matrix}
	\end{matrix}\right)
	\left(\begin{matrix}
		  \frac{1}{\sqrt{2}}\alpha_1\\
		  0\\
		  \frac{1}{\sqrt{2}}\alpha_2\\
		  \ 0\\
		       \frac{1}{\sqrt{2}}\alpha_1\\
		         0\\
		    \frac{1}{\sqrt{2}}\alpha_2\\
		   		 0
	\end{matrix}\right)=\frac{1}{2}
	\left(\begin{matrix}\beta_{1}\\ \bullet\\ \bullet\\ \bullet\\
		\beta_2\\ \bullet\\ \bullet \\ 
		\bullet
		\end{matrix}\right),
\end{equation}

\begin{figure}
	\centering
	\includegraphics[scale=0.6]{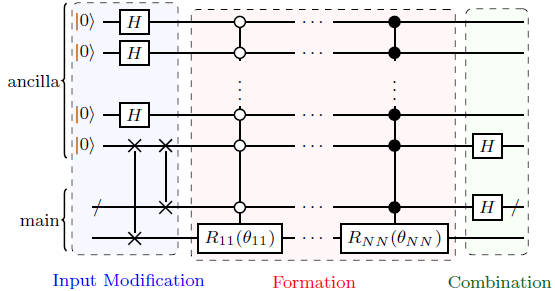}
	\caption{
General Universal circuit design \cite{Emmar2}: The angle values of the uniformly controlled rotation gates are directly determined from the matrix elements: $cos(\frac{\theta_{ij}}{2})=u_{ij}$. The matrix elements are tiled row by row on the diagonal of the matrix representation of the network formed by these rotation gates. The Hadamards
	at the end carry the same row elements from the diagonal to the  first rows of each $V_i$.
	The initial Hadamards and the SWAPs  modify the input. 
	}
	\label{fig:circuitfull}
\end{figure}

An instance circuit for this simulation technique is given in Fig.\ref{fig:circuitfull}
\cite{Emmar2}, which can simulate any given matrix based on Eq.(\ref{eqbm2}). The
separation of the main and the ancilla qubits in the circuit is largely artificial,
since their role can be interchanged in the circuit. To ease the verification process,
the circuit is divided into three blocks: \textit{Formation, Combination,} and
\textit{Input Modification} as shown in the figure. The verification of this circuit
is given in Appendix \ref{AppendixA}.

In the phase estimation process, for simplicity, we swap the states of the first $n$
ancilla qubits with the main qubits at the end of the circuit, and use the following
output instead of the one in Eq.(\ref{eqbm2}):
\begin{equation}
\left(\begin{matrix} 
		\kappa\beta_1\\ \vdots\\ 
		\kappa\beta_N\\ \bullet\\ 
	    \vdots \\ \bullet \end{matrix}\right),
\end{equation}
where the first $N$ states are chosen to be important states. \textbf{Measuring
all the ancilla qubits in the computational basis and post-selecting for the result
\ket{00...0}, one obtains in the remaining $n$ qubits in the state $\ket{\beta}$
(as introduced in Eq. (1)). In this case, the success probability $\kappa^2=1/2^{2n}$.}

\section{Finding Eigenvalues of Non-Unitary Matrices}
\label{SecEigen}
\begin{figure}
	\centering
	\includegraphics[scale=0.6]{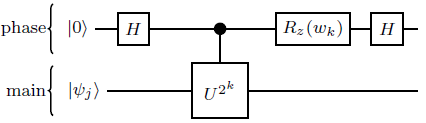}
	  \caption{
	 The iterative phase estimation algorithm for the $k$th iteration \cite{Miroslav,Abrams,Kitaev}.
	 In the circuit $|\psi_j\rangle$ is an eigenvalue of $U$, and the angle $w_k$ of the $R_z$
	 gate depends on the previously measured bits defined as $w_k=-2\pi(0.0x_kx_{k-1}...x_m)_{binary}$,
	 where $m$ is the number of digits determining the accuracy of the phase $\phi_j$. 
	 Note that $w_k$ is zero in the first iteration of the algorithm.}
	\label{fig:ipea}
\end{figure}
The  polar form of a complex eigenvalue, $\lambda_j$, belonging to a matrix, $U$,
can be written  as:
\begin{equation}
	\label{polarform}
	\lambda_j=r_j\ e^{-i2\pi \phi_j},
\end{equation}
where $\phi_j$ is the phase and $r_j=|\lambda_j|$ is the magnitude of the $j$th eigenvalue.
For unitary matrices,  $|\lambda_j|$ is equal to 1. Therefore,  the well-known efficient quantum phase estimation algorithm
(PEA) can be used to find the value of the phase, $\phi_j$,
and hence the eigenvalue $\lambda_j$. 
The circuit shown in  Fig.\ref{fig:ipea} is the iterative version of the phase estimation algorithm (IPEA) \cite{Miroslav,Abrams,Kitaev} which can also be used to estimate the value of $\phi_j$.
In each iteration of IPEA, a bit of the 
$m$-digit binary expansion of the phase is computed:
\begin{equation}
	\begin{split}
	\phi_j&= (0.x_1x_2...x_m)_{binary}
	\\&=x_12^{-1}+x_22^{-2}...\ x_{m-1}2^{-m+1}+x_m2^{-m},
	\end{split}
\end{equation}
where $m$ is the total number of iterations, which determines the precision. $x_k$ is the 
binary value found from the $k$th iteration of the IPEA using the $2^{m-k}$th power of
the matrix, $U^{2^{m-k}}$.

However, for non-unitary matrices, the phase estimation algorithm alone is not
enough. This is because two parameters, $\phi_j$ and $r_j$, need to be found to compute
the eigenvalue. In the following sections, we show that by using IPEA with the general circuit described in
the previous section, parameters $\phi_j$ and $r_j$ can be determined, and hence
eigenvalues of non-unitary matrices can also be found by the phase estimation
algorithm without additional cost. Since our circuit design  has a fixed size and
scheme, the application of IPEA to the circuit design also has a fixed design. Therefore,
we only need to set the angle values in each iteration of IPEA.
\begin{figure}
	\centering
	\includegraphics[scale=0.5]{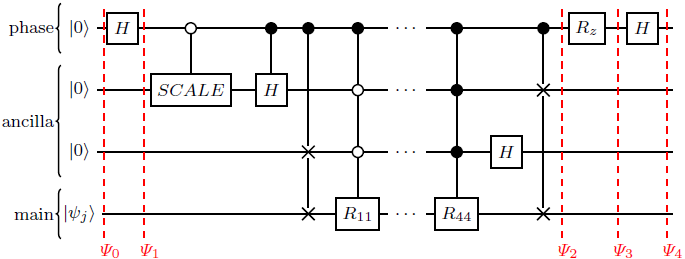}
	\caption{
	The iterative phase estimation algorithm with the circuit shown for a $2 \times 2$ matrix.
	  }
	\label{FigIPEA}
\end{figure}

\subsection{Estimation of $\phi_j$}
To use the general circuit design shown in Fig.\ref{fig:circuitfull}, 
within the IPEA framework, we need to determine the relationship between the chosen states
and the phase qubit resulting in the measurement. 
As described in the previous section,
the circuit scales the expected output of the chosen states by $\kappa$, which clearly
originates from the Hadamard gates. Since there are $2n$ Hadamard gates in the circuit, $\kappa=1/N$. 
In phase estimation, since all of the gates are
controlled by the upper qubit, this scaling effect of $\kappa$ exists only when the
phase qubit is one.  In order to have the same scaling  when also the phase qubit is  zero,
we allow all Hadamard gates on the ancilla qubits except the one on the first ancilla qubit  to operate without any control qubit. Instead of that controlled-Hadamard gate on the ancilla qubit, a scaling gate is used, which operates when the phase qubit is zero and has the following structure:
\begin{equation}
	SCALE=\left(\begin{matrix}
			 \frac{1}{\sqrt{N}}& \sqrt{1-\frac{1}{\sqrt{N}}}\\
			 -\sqrt{1-\frac{1}{\sqrt{N}}}& \frac{1}{\sqrt{N}}\\
		\end{matrix}\right)
\end{equation}
 The resulting circuit is shown for two qubit systems
in Fig.\ref{FigIPEA}. Here, we control all operations except the Hadamard  gates on the ancilla  qubits by the phase qubit. If the phase qubit is zero, for the input $\ket{\alpha}$ given in Eq.(\ref{eqbm}), this circuit produces the following on the remaining qubits:
\begin{equation}
	 \left(SCALE\otimes H^{\otimes n}\otimes I^{\otimes n}\right)\times 
	 \left(\begin{matrix}
	 	\alpha_1\\  \vdots\\ \alpha_N\\ 0\\
	    \vdots\\ 0
	    \end{matrix}\right)
	    =
	\left(\begin{matrix} 
		\kappa\alpha_1\\ \vdots\\ \kappa\alpha_N\\ \bullet\\ 
	    \vdots \\ \bullet \end{matrix}\right),
\end{equation}
where $\kappa=1/N$.
Therefore, we can estimate the phase  on the selected $N$ states, where the ancilla is zero in the output. 

Considering the circuit in Fig.\ref{FigIPEA}, in one iteration of the phase estimation algorithm, the effect of the eigenvalue,
represented in polar form, can be observed in  the evolution of the system as follows: 
\\
Fig.\ref{FigIPEA}, is:
\begin{equation}
	\Psi_0=\ket{0}_p\ket{00}_a|\psi_j\rangle,
\end{equation}
where $\ket{0}_p$, $\ket{00}_a$, and $|\psi_j\rangle$ represents the phase, ancilla and main qubits, respectively. $|\psi_j\rangle$ is the eigenstate of the operator $U$. After the first Hadamard gate on the phase qubit, the state becomes:
\begin{equation}
	\Psi_1=\frac{1}{\sqrt{2}}\left(\ket{0}_p\ket{00}_a
	|\psi_j\rangle 
	+\ket{1}_p\ket{00}_a\ket{\psi_j}\right)
\end{equation}
When the phase qubit is one, the application of the universal circuit generates the output $\kappa \lambda_j\ket{\psi_j}=\lambda_j/2\ket{\psi_j}$ on the chosen states. Since the Hadamard gates are not controlled, they are the only gates which operates when the phase qubit is zero.
Therefore, after the application of the universal circuit \textbf{with the
$SCALE$ gate (which scales the input state when the phase qubit is zero)},
we have the following:
\begin{equation}
\Psi_2=\frac{1}{\sqrt{2}}\left(\ket{0}_p\left(\frac{\ket{0}+\ket{1}}{\sqrt{2}}\otimes\frac{\ket{0}+\ket{1}}{\sqrt{2}}\right)_a
|\psi_j\rangle 
+\ket{1}_p\tilde{U}\ket{00}_a\ket{\psi_j}\right)
\end{equation}
where $\tilde{U}$ represents the universal circuit in Fig.\ref{fig:circuitfull}.      From the eigenvalue equation,  $U|\psi_j\rangle=\lambda_j|\psi_j\rangle$. On the chosen states, $\tilde{U}$ simulates $U$. Hence, if we consider chosen states (the states where the ancilla is $\ket{00}_a$), 
     $\tilde{U}|00\rangle_a|\psi_j\rangle=\kappa \lambda_j|00\rangle_a |\psi_j\rangle$ where $\kappa$ is the normalization constant. 
Therefore,  we have the following wave function describing
the chosen states in the first iteration:
\begin{equation}
\label{eq:state1}
\Psi_2=\frac {1}{2\sqrt{2}}\left(\ket{0}_p\ket{00}_a|\psi_j\rangle + 
e^{i2\pi\phi_j}|\lambda_j|\ket{1}_p\ket{00}_a|\psi_j\rangle\right),
\end{equation}
where the states which are not including $\ket{00}_a $ are ignored.   Application of the rotation gate around the z-axis on the phase qubit in the 
circuit guarantees that only one bit, $x_k$ in $0.x_k$, of the phase  
is estimated in each iteration. Hence, after this gate, the term $e^{i2\pi\phi_j}$ becomes  $e^{i2\pi(0.x_k)}$:
\begin{equation}
\label{eq:state1}
\Psi_3=\frac {1}{2\sqrt{2}}\left(\ket{0}_p\ket{00}_a|\psi_j\rangle + 
e^{i2\pi(0.x_k)}|\lambda_j|\ket{1}_p\ket{00}_a|\psi_j\rangle\right),
\end{equation}
 After the last Hadamard applied to the phase qubit,
the final state is as follows:
\begin{equation}
\Psi_4=\frac{1}{2\sqrt{2}}\left(\frac{|0\rangle+|1\rangle}{\sqrt{2}} |\psi_j\rangle+e^{i2\pi(0.x_k)}|\lambda_j|\frac{|0\rangle-|1\rangle}{\sqrt{2}}|\psi_j\rangle\right)
\end{equation}
or more concisely:
\begin{equation}
\label{phase}
\Psi_4=\frac{1}{4}\left(\left(1+e^{i2\pi(0.x_k)}|\lambda_j|\right)|0\rangle+\left(1-e^{i2\pi(0.x_k)}|\lambda_j| \right) |1\rangle\right)|\psi_j\rangle.
\end{equation}
Here, because of the system size, $\kappa=1/2$. For general case,  $\kappa$ is directly related to the size of the ancilla and is $1/N$ if there are $(n+1)$ qubits in the ancilla. Thus, the above equation can be represented in general form as follows:
\begin{equation}
\label{phase}
\Psi_4=\frac{\kappa}{2}\left(\left(1+e^{i2\pi(0.x_k)}|\lambda_j| \right) 
|0\rangle+\left(1-e^{i2\pi(0.x_k)}|\lambda_j| \right) |1\rangle \right)|\psi_j\rangle.
\end{equation}
Since $x_k$ is a bit value, it can be either 0 or 1.
When $x_k=0$, since $e^{i2\pi0}=1$, the above equation reduces to:
\begin{equation}
\Psi_4=\frac{\kappa}{2}\left(\left(1+|\lambda_j|\right)|0\rangle+ \left(1-|\lambda_j| \right)|1\rangle \right)|\psi_j\rangle.
\end{equation}
In the case, when $x_k=1$, since $e^{i\pi}=-1$,  Eq.(\ref{phase}) becomes:
\begin{equation}
\Psi_4=\frac{\kappa}{2}\left( \left(1-|\lambda_j|\right)|0\rangle+ \left(1+|\lambda_j|\right)|1\rangle \right)|\psi_j\rangle.
\end{equation}
Since $\frac{\kappa}{2}\left(1-|\lambda_j| \right)\le \frac{\kappa}{2}\left(1+|\lambda_j|\right)$;  if $x_k=0$, the probability to find the phase
qubit in $\ket{0}$ state is higher  then the probability to find it in $\ket{1}$.  
If $x_k=1$, then the probability to find the phase qubit in \ket{1} is  higher. Thus, from the measurement
on the phase qubit  $x_k$ can be determined.  

In each iteration of the phase estimation algorithm,
a different bit of the phase is determined in this manner. For the $k$th iteration of the
phase estimation algorithm to determine the $k$th digit of the phase, we
see $\frac{\kappa}{2}(1-|\lambda_j|^{2^k})\le \frac{\kappa}{2}(1+|\lambda_j|^{2^k})$.
Hence,  for $|\lambda_j|<1$, the more we increase the power of $U$ or the accuracy,
the more the amplitudes of the states, $|0\rangle$ and $|1\rangle$,  come closer. 
\textbf{In this case, a single iteration of the PEA may repeat the protocol
several times to collect enough statistics to determine the phase bit with high confidence. 
This is particularly important for later iterations, where, for $|\lambda_j|<1$,
the probability of finding the phase qubit in \ket{0} or \ket{1} is exponentially
(in the iteration index $k$) close.
}
\subsection{Computing $|\lambda_j|$}
As we have shown above, the output probability of the phase qubit when the ancilla is
zero is determined by $\frac{\kappa}{2}(1+|\lambda_j|)$. We measure $|0\rangle$ or
$|1\rangle$ with probability $P$  equal to $\frac{\kappa}{2}(1+|\lambda_j|)$. Hence, we get 
\begin{equation}
\label{eq:probability}
\begin{split}
&P=(\frac{\kappa}{2}(1+|\lambda_j|))^2,\\
&|\lambda_j|=\frac{2\sqrt{P}}{\kappa}-1.
\end{split}
\end{equation}
Since  $\kappa$ and $P$ are known (for a $2 \times 2$ matrix, the value of $\kappa$ is
$\frac{1}{2}$ due to two Hadamard gates in the circuit), $|\lambda_j|$ can be determined from the statistics of the measurement. Therefore, the accuracy of the estimate value of
$|\lambda_j|$ can be  further improved by using all iterations of the phase estimation
algorithm: For the $k$th iteration, as a general form, the following is obtained:
\begin{equation}
\begin{split}
&P_{(k)}=\left(\frac{\kappa}{2}(1+|\lambda_j|^{2^k})\right)^2,\\
&|\lambda_j|^{2^k}=\frac{2\sqrt{P_{(k)}}}{\kappa}-1.
\end{split}
\end{equation}
Taking the average of the estimates from different phase estimation iterations,
the estimate of $|\lambda_j|$ may become more accurate. Finally, by using values of
$|\lambda_j|$ and $\phi_j$ in Eq. (\ref{polarform}), we can compute the eigenvalue
of the non-unitary matrix.

\section{Discussion}
In the phase estimation algorithm,  to be able to generate the matrix elements by using rotation gates around y- and z-axis,  the absolute values of the elements should be less than one. 
One way to achieve this is to divide all the elements by the absolute value of the maximum element. 
 Based on the norm and the eigenvalue relationship, this guarantees that the absolute value of the largest eigenvalue is less than the absolute row or column sums of the matrix which can be maximum $N$ (i.e. because the absolute value of the maximum is 1.). The eigenvalue is not required to be less than one. However, if the eigenvalue is greater than 1, this approach may require scaling in the very iteration of the algorithm since in the powers there is a possibility the elements may become greater than one again. 
Another approach to scale the elements is to use the 1-norm or the infinity-norm of the matrix, which are easy to compute. Scaling the matrix by the norm makes the eigenvalue less than 1, and so in the power of the matrix the elements goes to zero. 
Hence, the scaling is only done at the beginning.
 Therefore, in the $k$th iteration of the phase estimation algorithm, we have $(\lambda_j^{2^k}/\mu^{2^k}<0)$, where $\mu$ is the scaling used at the beginning.
  Hence, if $|\lambda_j|<1$, without any scaling, the more we increase the power of $U$  or the accuracy,
the more the amplitudes of the states, $|0\rangle$ and $|1\rangle$,  come closer.
 For $|\lambda_j|\ll 1$, after a few iterations, it becomes infeasible to distinguish whether the phase qubit is 0 or 1. Scaling the powers of $U$ which generates the eigenvalue $\lambda_j^{2^k}/\mu$ can be used to remedy this problem.

The accuracy for determining the value of $r_j$, requires to determine the terms $(1+r_j)$ and $(1-r_j)$ in the probabilities for finding the phase qubit in \ket{0} or \ket{1} from the statistics on the measurements. The accuracy of in these statics differs based on the measurement protocol, the quantum state, the underlying quantum machine, and even the amount of the entanglement \cite{Usami2003,Bogdanov2011}. 

\subsection{Algorithmic Complexity}
The complexity of one-iteration of the phase estimation algorithm is mainly dominated by the complexity of implementing the given operator. There is also need to obtain the state tomography of the phase qubit in order to determine the absolute value of the eigenvalue which differs based on the underlying quantum machine used for the phase estimation algorithm. 

It is known that the complexity of implementing an operator on quantum computer requires $O(N^2)$ number of one and two qubit operations \cite{Chuang}. 
In addition, more efficient circuits are possible for the sparse matrices. The general circuit described in Fig.\ref{fig:circuitfull} also obeys these complexity behavior (see ref.\cite{Emmar2} for the detailed complexity analysis of the circuit.). 
In the circuit, there is a quantum network composed of $N^2$ rotation gates uniformly controlled by the first $2n$ qubits. The decomposition of this network requires $2^{2n}$ number of CNOT and $2^{2n}$ single rotation operations, which is explained in Appendix \ref{AppendixB}. Therefore, the circuit in total requires $2^{2n}$ CNOT, $2^{2n}$ single rotation, $2n$ Hadamard, and $n$ SWAP gates \cite{Emmar2}. Hence, the total complexity is $O(N^2)$ .  
Also note that for zero elements, there is no need to use rotation gates. Thus, for the sparse matrices, the number of operations can be reduced \cite{Emmar2}. 

Since the complexity of an iteration of the phase estimation algorithm is dominated by the complexity of the implementing the given operator, each iteration of the algorithm requires $O(N^2)$ operations. Then for $m$ number of iterations, we get $O(mN^2)$ as the complexity. 

When the powers of the matrix cannot be efficiently computed, the complexity of computing the power of matrix $U$ on classical computers
should also be  considered. The power of a matrix can be found   by using  the successive squaring method which requires only one matrix multiplication. 
Then the total complexity  becomes the combination of these two: 
\begin{equation}
\label{Eq:complexity}
O(mM)+O(mN^2)=O(m(N+M)),
\end{equation} 
where $m$ is the number of iterations, and $M$ is the complexity of
matrix multiplication. Note that even if the powers of the matrix needs to be computed, this is still at least as efficient as the classical algorithms which requires in general $O(N^3)$ time for finding an eigenvalue \cite{Golub}.
\textbf{Also, note that the algorithmic complexity in Eq.(\ref{Eq:complexity}) does
not include the complexities of measuring and determining the value of $r_j$. These are
dependent on the measurement protocol, the quantum state, and the underlying quantum
machine. In addition, the total number of iterations $m$ may be large in cases
where, for $|\lambda_j|<1$, the probability of finding the phase qubit in
\ket{0} or \ket{1} states gets exponentially (in the iteration index $k$) close,
since a single iteration of the PEA may require repeating the protocol several
times to collect enough statistics to determine the phase bit with high confidence.}

In comparison to the work done by Wang et al. \cite{Hefeng-MPEA}, the success probability decreases exponentially in both algorithms described in this paper and their paper. However, in our case by scaling the matrix elements in every iteration of the algorithm, the probability can  be increased. Furthermore, here, one can finds the angle values for the rotation gates very easily (see Appendix \ref{AppendixB1}) to implement the given operator. However, in their method, one needs to find subsystems $H_A$ and $H_B$ and their interaction $H_{AB}$ in order to implement the given $H$: $H=H_A+H_B+H_{AB}$.  In terms of quantum complexity, since their method is based on the non-unitary successive projective measurements on $H_A$  which cannot be implemented deterministically, the complexity is exponentially gets larger \cite{Hefeng-MPEA}.  On the other hand, in their case finding the powers of the given operator is easier, however, when the size of the subsystem $H_A$ is large, the complexity is dominated by the complexity of the measurements.   

\section{Application to Non-Hermitian Quantum Systems}
Some problems can be extremely hard or even impossible to solve within the
framework of the standard formalism of quantum mechanics where the observable properties of the dynamic nature are real and associated with the eigenvalues of Hermitian operators. Most extensions of the standard Hermitian formalism of quantum mechanics are 
equivalent to use a non-Hermitian operator instead of a Hermitian one. Resonance phenomena, where the particles are temporarily trapped by the potential, can be explored within  the non-Hermitian quantum mechanics. In the non-Hermitian
quantum mechanics, the lifetime of the resonant states are
proportionally dependent on the behavior of the imaginary part of the eigenvalue 
\cite{Pablo,Moiseyev2011}. 
Suppose we have the non-unitary operator $U=e^{-i\mathcal{H}t/\hbar}$ for a non-Hermitian
Hamiltonian $\mathcal{H}$ with energy eigenstates  $|\psi_j\rangle$ and corresponding
energy eigenvalues $E_j$, i.e., $\mathcal{H}|\psi_j\rangle=E_j|\psi_j\rangle$. Since $E_j$
is an eigenvalue of $\mathcal{H}$; if $t$ and $\hbar$ are set to $1$, then $e^{-iE_j}$
is the eigenvalue of $U$. Therefore, a $N \times N$ non-unitary transformation $U$ has
eigenvectors $|\psi_1\rangle$, $|\psi_2\rangle$, ..., $|\psi_N\rangle$,
with corresponding eigenvalues $\lambda_j=|\lambda_j|e^{-i2\pi \phi_j}$. As shown in
Sec.\ref{SecEigen}, the eigenvalue $\lambda_j$ of the matrix $U$ can be estimated. This
also allows us to determine the corresponding eigenvalue $E_j$ of the Hamiltonian
$\mathcal{H}$ from $\lambda_j=e^{-iE_j}$.
Note that in our phase estimation procedure, since the matrix elements are directly used in the circuit, one can also directly use the Hamiltonian matrix, $\mathcal{H}$, instead of $U$.
\subsection{Example}
As an example, we  consider  the following radial
 Hamiltonian  \cite{Pablo}:
\begin{equation}
\mathcal{H}=
-\frac{1}{2}\frac{d}{dr}(r^2\frac{d}{dr})+(\frac{r^2}{2}-J)e^{-ar^2},
\end{equation}
where  $J$ and $a$ are free potential parameters ($J$ is the depth of the potential in the potential graph. $a$ is related to the width of the potential barrier.)  The potential term, $(\frac{r^2}{2}-J)e^{-ar^2}$, in the above Hamiltonian exhibits
 predissociation resonances which are quasibound
states associated with the complex part of the eigenvalues \cite{Pablo}.  For the illustration purpose, using
only  two basis functions (the orthonormalized eigenfunctions
of the harmonic oscillator, where mass equals one and the frequency equals $a$) and setting $J=0.1$ and $a=0.1$;
the Hamiltonian matrix is found as \footnote{This matrix is given by Pablo Serra,
private communication, 2012. \cite{Pablo}}:
\begin{equation}
H=\left(\begin{matrix}
1.4216 - 0.1576i & 0.2782 + 0.2802i\\
0.2782 + 0.2802i &0.6807 - 0.2361i
\end{matrix}\right),
\end{equation}
which has the eigenvector:
\begin{equation}
|\psi_j\rangle =
\left(\begin{matrix}
-0.3790 - 0.1962i\\
0.9044 \end{matrix}\right),
\end{equation} 
and the ground state eigenenergy $( 0.6249 - 0.4139i)$.
Since the above Hamiltonian matrix is non-Hermitian,  the time evolution operator found from $U=e^{-i\mathcal{H}t/\hbar}$ is non-unitary, which is found as:
\begin{equation}
\begin{split}
U&=
e^{i\mathcal{H}}\\&=\left(\begin{matrix}
0.2588 + 1.1214i& -0.4569 - 0.1109i\\
-0.4569 - 0.1109i & 1.0594 + 0.7394i
\end{matrix}\right),
\end{split}
\end{equation} 
with eigenvalue $1.2268 + 0.8849i$.
We have used the phase estimation method described above to simulate this non-unitary operator and
determine the complex eigenvalues of $\mathcal{H}$. 
In the simulation, since the absolute values of the matrix elements of $U$ are greater than 1, we scale the elements by the 1-norm of the matrix. This makes the matrix elements of $U$  go to zero in the limit. We use 11 iterations in the phase estimation algorithm. After the 11th iteration, the probability difference between seeing the phase qubit 0 or 1 becomes almost equal.

 Therefore, from the simulation output, we find the eigenvalue of $U$ as $1.22255 + 0.88355i$,
and so the eigenvalue of $\mathcal{H}$ as $0.6259 - 0.4089i$, which gives an error
in the order of $10^{-3}$. This comes from computer round-off errors and the limited available
powers of $U$ (when the power of $\lambda$ goes to zero, we can no longer distinguish
between $|0\rangle$ and $|1\rangle$ on the phase qubit; the probabilities becomes
the same). The simulation details are  given
in  Appendix \ref{AppendixB}. 

\section{Conclusion}
We have presented  a general scheme for the execution of the phase estimation algorithm
for non-unitary matrices. Since the circuit design  used for non-unitary matrices is a
general programmable circuit, the circuit for the phase estimation algorithm is also
universal: It can be used for any type of matrix to compute its eigenvalues. 
We have shown that the complex eigenvalues of non-Hermitian quantum systems can be found
by using this method. As an example we have shown how to explore the  resonance states of a model
Hamiltonian.  
\section{Acknowledgement}
We thank Pablo Serra for providing the Hamiltonian matrix used as an example in this paper.
\textbf{We also thank the two anonymous referees for their help in improving the clarity
of this paper.}
This work is supported by the NSF Centers for Chemical Innovation: Quantum Information
for Quantum Chemistry, CHE-1037992.
\appendix{
\section{Verification of the Circuit in Fig.\ref{fig:circuitfull}}
\label{AppendixA}
 The circuit is divided in three blocks: \textit{Formation, Combination,} and \textit{Input Modification} as shown in the figure.
 In the \textit{Input Modification} block, the initial input is modified to $|\varphi\rangle$. Then, the matrix $V$ is constructed in the \textit{Formation} and \textit{Combination} blocks. 
\subsubsection{Formation Block:} 
The second block in Fig.\ref{fig:circuitfull} is called \textit{Formation Block} which consists of uniformly controlled rotation gates for each element of $U$. For the matrix element $u_{ij}$, assumed to be real, the rotation gate $R_{ij}(\theta_{ij})$ is defined as follows:
\begin{equation}
	R_{ij}(\theta_{ij})=
	\left(\begin{matrix}
		\cos(\frac{\theta_{ij}}{2}) &\sin(\frac{\theta_{ij}}{2})\\
		-\sin(\frac{\theta_{ij}}{2})&\cos(\frac{\theta_{ij}}{2})
	\end{matrix}\right)
\end{equation}
where $\theta_{ij}$ is determined from the value of the element in accordance with the equality
$cos(\frac{\theta_{ij}}{2})=u_{ij}$  to get the following:
\begin{equation}
	R_{ij}(\theta_{ij})=
	\left(\begin{matrix}
		u_{ij} &\sqrt{1-u_{ij}}\\
		-\sqrt{1-u_{ij}}&u_{ij}
	\end{matrix}\right)
\end{equation}
Based on the linear indices of the elements (row-wise), $R_{ij}$s are controlled by the different states of $(2n)$ qubits: e.g., $u_{12}$ is the second element in the matrix and has a linear index of $2$. Hence, the control of $R_{12}$ is set accordingly so that this gate operates when the first $(2n)$ qubits in $\ket{0\dots01}$ state. Combination of these  controlled rotation gates forms a network as the second block in Fig.\ref{fig:circuitfull}. This block has the following matrix representation in the computational basis: 
\begin{equation}
	\label{Eqmat1}
	F=\left(
	\begin{matrix}
		R_{11}&\\
		&R_{12}&\\
		&&\ddots&\\
		&&&R_{NN} 
	\end{matrix}\right),\ 
	R_{ij}=
	\left(\begin{matrix}
		u_{ij} &\sqrt{1-u_{ij}}\\
		-\sqrt{1-u_{ij}}&u_{ij}
	\end{matrix}\right).
\end{equation}
Here, we have used rotations around the y-axis since the matrix elements are assumed to be real. However,
for the complex elements of $U$, a product of $R_z$ and $R_y$ gates is needed for
each $R_{ij}$. For instance, if $u_{ij} = e^{i\phi}cos(\theta)$, $e^{i\phi}$ is created by $R_z$, and $cos(\theta)$ is by $R_y$.
\subsubsection{Combination Block:}
For a system
of $(2n+1)$ qubits, the third block in Fig.\ref{fig:circuitfull}  is defined as  the application of the Hadamard gates to 
$(n+1)$th, $(n+2)$th, ..., $(2n-1)$th, and $(2n)$th qubits from the top of the circuit: i.e. $(I^{\otimes n}\otimes H^{\otimes n}\otimes I)$, where $H$ and $I$ are the Hadamard and identity matrices, respectively. The matrix representation of this block is as follows:

\begin{equation}
	C=\left(\begin{matrix}
		C_{block}\\
		&C_{block}\\
		&&\ddots\\
		&&&C_{block}\\
	\end{matrix}\right), \text{where\ }
	C_{block}=\label {eqH}
	\left(\begin{matrix}
		k&0&\dots& k &0\\
		0&k&\dots&0&k\\
		\vdots&\vdots&\vdots&\vdots&\vdots\\
		k&0&\dots&k&0\\
		0&k&\dots&0&k\\
	\end{matrix}\right)_{2N\times2N}.
\end{equation}
Note that $C_{block}$  also has negative elements, however, they are not located in the first row. Therefore, they shall not affect the predetermined states.
The application of the above matrix $C$ to the  matrix $F$ defined in Eq.(\ref{Eqmat1}) forms 
the matrix $V$ defined in Eq.(\ref{eqbm2}) where the same row elements of $U$ are located on the leading rows of each
$V_i$ ($i$ represents the row index of $U$):
\begin{equation}
	\label{Eqmat2}
	\begin{split}
	V&=CF\\
	&=
	\left(\begin{matrix}
		C_{block}\\
		&C_{block}\\
		&&\ddots\\
		&&&C_{block}\\
	\end{matrix}\right)
	\left(\begin{matrix}
		\begin{matrix}
		u_{11}&\sqrt{1-u_{11}}\\-\sqrt{1-u_{11}}&u_{11}
		\end{matrix}&\\
		&\ddots&\\
		&&\begin{matrix}
		u_{NN}&\sqrt{1-u_{NN}}\\-\sqrt{1-u_{NN}}&u_{NN}
		\end{matrix}
	\end{matrix}
	\right)
	\\&=
	\left(\begin{matrix}
		\begin{matrix}
		ku_{11}&\bullet&ku_{12}&\dots &\bullet& ku_{1N}&\bullet\\
		\vdots&\vdots&\vdots&&\vdots&\vdots&\vdots
		\end{matrix}& &
		\\ 
		& \ddots &
		\\
		& &\begin{matrix}
		ku_{N1}&\bullet&ku_{N2}&\dots &\bullet& ku_{NN}&\bullet	\\
		\vdots&\vdots&\vdots&&\vdots&\vdots&\vdots
		\end{matrix}
	\end{matrix}\right)
	.
	\end{split}
\end{equation}
Since the negative elements are not in the first row of $C_{block}$, the resulting leading rows are not affected by these elements. The elements represented by the symbol \lq\lq{}$\bullet$\rq\rq{} are disregarded by modifying the input to the circuit in the modification block explained below.
\subsubsection{Input Modification Block:}
The initial input
to the circuit in Fig.\ref{fig:circuitfull} is defined as 
$\tilde{\ket\alpha}=\ket{0..0}\otimes\ket{\alpha}$ where $\ket{0...0}$
is the input to the ancilla qubits. The first block in Fig.\ref{fig:circuitfull} consists of
 the Hadamard gates on the first $n$ qubits
and sequential swap operations between the $(n+1)$th and the remaining
last $n$ qubits\textbf{ (We apply swap operations between the ancilla $(n+1)$th
qubit and the main qubits: First we swap the $(n+1)$th qubit with the $(2n+1)$th,
then the $(n+1)$th with the $(2n)$th, then the $(n+1)$th with the $(2n-1)$th,
and so on. Finally we swap the $(n+1)$th with the $(n+2)$th)}. This block modifies
the input $\tilde{\ket\alpha}$ in a way that in the application of $V=CF$ to
the input, the output is not affected by the elements represented by
\lq\lq{}$\bullet$\rq\rq{} between $ku_{ij}$ and $ku_{i(j+1)}$. Hence, the application
of this block to the initial input transforms $\tilde{\ket{\alpha}}$ to
$\ket\varphi$ in Eq.(\ref{eqbm2}):
\begin{equation}
	\label{eq5}
	  \begin{split}
	    \tilde{\ket{\alpha}} \rightarrow& \ket\varphi
	    \\ [\alpha_1\ \alpha_2\ \dots \alpha_N\ 0
	    \dots 0]^T \rightarrow&
	    [\gamma\alpha_1\ 0\ \gamma\alpha_2\dots 0\ \gamma\alpha_N\dots\
	    \gamma\alpha_1\ 0\ \gamma\alpha_2\ \dots\ 0\ \gamma\alpha_N\ 0]^T,
	  \end{split}
\end{equation}
where $\gamma$ is a normalization constant.

Consequently, the circuit in Fig.\ref{fig:circuitfull} describes the operation $V\ket\varphi=CF\ket\varphi$ which simulates any matrix $U$, having
elements less than or equal to 1, on the following normalized set of $N$ states:
$\{\ket{0\dots000}_a\ket{0\dots0}$,
$\dots$, $\ket{0\dots110}_a\ket{0\dots0}\}$, where $\ket{\dots}_a$ represents the ancilla qubits. This is shown in Eq.(\ref{eqbm2}). 

\textbf{For illustration purposes, we also present the full forms of  the operators
for the formation, combination and input modification blocks and the output vector
for the simulation of the following $2 \times 2$ arbitrary matrix \cite{Emmar2}:}
\begin{equation}
U=
	\left(\begin{matrix}
	u_{11}&u_{12}\\
	u_{21}&u_{22}\\\end{matrix}\right)
\end{equation}

The full form of the matrix for the formation block is as follows:
\begin{equation} F= \left(
\begin{array}{cccccccc}
 u_{11} & \sqrt{1-u_{11}^2} & 0 & 0 & 0 & 0 & 0 & 0 \\
 -\sqrt{1-u_{11}^2} & u_{11} & 0 & 0 & 0 & 0 & 0 & 0 \\
 0 & 0 & u_{12} & \sqrt{1-u_{12}^2} & 0 & 0 & 0 & 0 \\
 0 & 0 & -\sqrt{1-u_{12}^2} & u_{12} & 0 & 0 & 0 & 0 \\
 0 & 0 & 0 & 0 & u_{21} & \sqrt{1-u_{21}^2} & 0 & 0 \\
 0 & 0 & 0 & 0 & -\sqrt{1-u_{21}^2} & u_{21} & 0 & 0 \\
 0 & 0 & 0 & 0 & 0 & 0 & u_{22} & \sqrt{1-u_{22}^2} \\
 0 & 0 & 0 & 0 & 0 & 0 & -\sqrt{1-u_{22}^2} & u_{22}
\end{array}
\right)
\end{equation}

The combination matrix $C$ and the matrix for the input modification $M$ are defined as:
\begin{equation}
C = \left(
\begin{array}{cccccccc}
 \frac{1}{\sqrt{2}} & 0 & \frac{1}{\sqrt{2}} & 0 & 0 & 0 & 0 & 0 \\
 0 & \frac{1}{\sqrt{2}} & 0 & \frac{1}{\sqrt{2}} & 0 & 0 & 0 & 0 \\
 \frac{1}{\sqrt{2}} & 0 & -\frac{1}{\sqrt{2}} & 0 & 0 & 0 & 0 & 0 \\
 0 & \frac{1}{\sqrt{2}} & 0 & -\frac{1}{\sqrt{2}} & 0 & 0 & 0 & 0 \\
 0 & 0 & 0 & 0 & \frac{1}{\sqrt{2}} & 0 & \frac{1}{\sqrt{2}} & 0 \\
 0 & 0 & 0 & 0 & 0 & \frac{1}{\sqrt{2}} & 0 & \frac{1}{\sqrt{2}} \\
 0 & 0 & 0 & 0 & \frac{1}{\sqrt{2}} & 0 & -\frac{1}{\sqrt{2}} & 0 \\
 0 & 0 & 0 & 0 & 0 & \frac{1}{\sqrt{2}} & 0 & -\frac{1}{\sqrt{2}}
\end{array}
\right),\quad 
M = \left(
\begin{array}{cccccccc}
 \frac{1}{\sqrt{2}} & 0 & 0 & 0 & \frac{1}{\sqrt{2}} & 0 & 0 & 0 \\
 0 & 0 & \frac{1}{\sqrt{2}} & 0 & 0 & 0 & \frac{1}{\sqrt{2}} & 0 \\
 0 & \frac{1}{\sqrt{2}} & 0 & 0 & 0 & \frac{1}{\sqrt{2}} & 0 & 0 \\
 0 & 0 & 0 & \frac{1}{\sqrt{2}} & 0 & 0 & 0 & \frac{1}{\sqrt{2}} \\
 \frac{1}{\sqrt{2}} & 0 & 0 & 0 & -\frac{1}{\sqrt{2}} & 0 & 0 & 0 \\
 0 & 0 & \frac{1}{\sqrt{2}} & 0 & 0 & 0 & -\frac{1}{\sqrt{2}} & 0 \\
 0 & \frac{1}{\sqrt{2}} & 0 & 0 & 0 & -\frac{1}{\sqrt{2}} & 0 & 0 \\
 0 & 0 & 0 & \frac{1}{\sqrt{2}} & 0 & 0 & 0 & -\frac{1}{\sqrt{2}}
\end{array}
\right)
\end{equation}

For the initial input  $\ket{\bm0}\ket{\alpha}=\hat{\ket{\alpha}}$,
we find the following final state:

\begin{equation}
CFM\hat{\ket{\alpha}}=V\ket{\varphi}=\frac{1}{2}\left(\begin{matrix}
 \alpha _{1} u_{11}+\alpha _{2} u_{12} \\
 - \alpha _{1} \sqrt{1-u_{11}^2}- \alpha _{2} \sqrt{1-u_{12}^2} \\
 \alpha _{1} u_{11}-\alpha _{2} u_{12} \\
 -\alpha _{1} \sqrt{1-u_{11}^2}+ \alpha _{2} \sqrt{1-u_{12}^2} \\
 \alpha _{1} u_{21}+\alpha _{2} u_{22}\\
 - \alpha _{1} \sqrt{1-u_{21}^2}- \alpha _{2} \sqrt{1-u_{22}^2} \\
 \alpha _{1} u_{21}-\alpha _{2} u_{22} \\
 - \alpha _{1} \sqrt{1-u_{21}^2}+ \alpha _{2} \sqrt{1-u_{22}^2}
\end{matrix}\right)
\end{equation}

Clearly the normalized states $|000\rangle$ and $|100\rangle$ simulate the original given system.
\section{Simulation Details}
\label{AppendixB}
\subsection{The decomposition of a multi controlled network}
\label{AppendixB1}
The circuit in Fig.\ref{fig:circuitfull} includes a network of rotation gates in the formation block which dominates the complexity of the circuit.
The uniformly controlled networks such as the one in Fig.\ref{circuit3} controlled
by $k$ qubits can be decomposed in terms of $2^k$ CNOT gates and $2^k$ single
rotation gates\cite{Mottonen}. For instance, the circuit as illustrated for $k=2$
in Fig.\ref{circuit3} can be decomposed as in Fig.\ref{circuit4}.
\begin{figure}
  \subfloat[]{\centerline{
\includegraphics[width=3.5in]{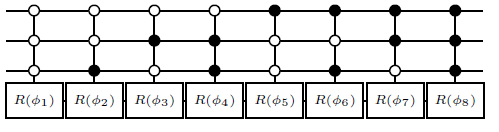}
  }\label{circuit3}
}

\subfloat[]{ \centerline{\resizebox{\linewidth}{!}{%
\includegraphics[width=3.5in]{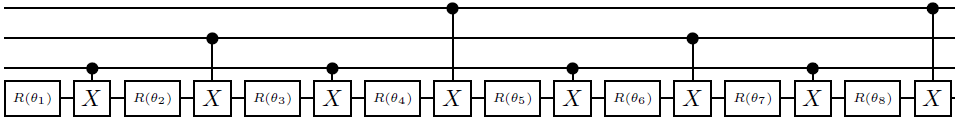}
  }}
\label{circuit4}}
  \caption{
(a) A uniformly controlled multi-qubit network. (b) The decomposition of the network in (a)
into CNOT and single quantum gates. The change of the bits in the gray code representations
determines the control qubit for the CNOT gates. The network in the phase estimation
algorithm includes two consecutive circuits such as in (b) with $R_z$ and $R_y$ single gates.
  }
\label{circuit3full}
\end{figure}
The angle values in the decomposed circuit are solutions of the system of the linear
equation $M^{k}\boldsymbol\theta=\boldsymbol\phi$:
\begin{equation}
M^{k}\left(\begin{array}{c}
\theta_1\\
\theta_2\\
\vdots\\
\theta_{2^{k}}
\end{array}\right)=\left(\begin{array}{c}
\phi_1\\
\phi_2\\
\vdots\\
\phi_{2^{k}}
\end{array}\right),
\end{equation} where $k$ is the number of control qubits in the network, and
the entries of $M$ are defined as:
\begin{equation} 
M_{ij}=(-1)^{b_{i-1}.g_{j-1}},
\end{equation}
in which the power term is found by taking the dot product of the standard binary
code of the index $i-1$, $b_{i-1}$, and the binary representation of $j-1$th Gray
coded integer, $g_{j-1}$. Since $M^k$ is a column permuted version of the Hadamard
matrix, we see that $M$ is unitary. Thus, $(M^k)^{-1}=2^{-k}(M^k)^T$, and the new angle
values in the decomposed circuit are results of the matrix vector multiplication\cite{Mottonen}: 
\begin{equation}
\label{matvec}
\boldsymbol\theta=2^{-k}(M^k)^{T}\boldsymbol\phi.
\end{equation}
\subsection{Simulation Details for the Example System}
\label{AppendixB2}
In Fig.\ref{FigIPEA}, we have a multi controlled network composed of 4 gates. 
This network basically comes from the \textit{Formation} step of the circuit design
method, which has been represented in matrix form as $F$ in Eq.(\ref{Eqmat1}).
Since the elements of $U$ are complex, we need to have rotations around the $z-$axis
and the $y-$axis. Hence, the above matrix is the product of two matrices $F_z$
for rotation around the z-axis, and $F_y$ for rotations around the y-axis: $F=F_zF_y$:
\begin{equation}
F=\left(
\begin{matrix}
R_{11}^z&\\
&\ddots&\\
&&R_{NN}^z 
\end{matrix}
\right)
\left(
\begin{matrix}
R_{11}^y&\\
&\ddots&\\
&&R_{NN}^y 
\end{matrix}
\right)
\end{equation}
To complete this network to a whole uniformly controlled network, we assume
that the initial four other gates are identity. Hence, the final decomposition
for $2 \times 2$ matrix, the decomposed circuit includes: 4 Hadamard gates, 16 CNOT
and 16 single gates (8 CNOTs and 8 $R_y$ gates for the $F_y$; and 8 CNOTs and 8 $R_z$
gates for the $F_z$), and 2 swaps. 

The parameters for the last iteration (The operator is $U^{2^0}$.) of the phase estimation algorithm is shown below, where after scaling the elements of $U$ by the matrix norm $||U||_1$ (maximum of the absolute sums of the columns), we find the angle values  for $R_y$ and $R_z$ gates: 
\begin{center}
\begin{tabular}{|c|c|c|c|c}
\hline
	Matrix Elements	&	Scaled Elements	&	Angles for $R_y$s	&	Angles for $R_z$s	\\
\hline	   0.2588 + 1.1214i	&	   0.1469 + 0.6364i	&	1.0521	&	-0.9634	
\\	  -0.4569 - 0.1109i	&	  -0.2593 - 0.0629i	&	0.0278	&	0.1837	
\\	  -0.4569 - 0.1109i	&	  -0.2593 - 0.0629i	&	-0.2486	&	1.9401	
\\	   1.0594 + 0.7394i	&	   0.6012 + 0.4196i	&	0.0278	&	0.1837	
\\		&		&	-0.0278	&	-0.1837	
\\		&		&	0.2486	&	-1.9401	
\\		&		&	-0.0278	&	-0.1837	
\\		&		&	-1.0521	&	0.9634	\\
\hline
\end{tabular}
\end{center}
The angle value for the $R_z$ gate on the first qubit is -2.51572849.
The output of the phase estimation algorithm on the chosen states and the normalized probability of the phase qubit are shown below.
\begin{center}
\begin{tabular}{|l|>{\centering\arraybackslash}m{3cm}|>{\centering\arraybackslash}m{3cm}|}
\hline
States &	Probabilities on the Chosen States &	Normalized Probabilities of the phase qubit\\
\hline 	0	&	0.0002	&	0.0058
\\	1	&		0.001	&	0.9942
\\	2	&		0.0393	&	
\\	3	&			0.1765	&	\\
\hline
\end{tabular}
\end{center}

 The bit values of the phase is found as $(11100110100)$ which corresponds to 0.9004. 
The absolute value of the eigenvalue is found from the ratio of the probability values of the phase qubit:
\begin{equation}
\frac{(1+\frac{|\lambda|}{||U||_1})^2}{(1-\frac{|\lambda|}{||U||_1})^2}=\frac{0.9942}{0.0058}
\end{equation}
From the above, we find $|\lambda|=0.8560\times||U||_1=1.5084$.  The eigenvalue of $U$ is found as $e^{i2\pi 0.9004}\times1.5084=1.22255 + 0.88355i$. The eigenvalue of the Hamitlonian matrix is found from $(log(1.22255 + 0.88355i)/i)$ which is $0.62581 - 0.41105i$.

}
\bibliographystyle{spphys}
\bibliography{bibtexF}

\end{document}